\documentclass[showpacs,aps,prl,twocolumn,superscriptaddress]{revtex4} 
\usepackage[pdftex]{graphicx} 
\usepackage{amsmath} 
\usepackage{amsfonts} 
\usepackage{amssymb} 
\usepackage{amsbsy} 
\usepackage{ulem} 
\usepackage{color} 
\usepackage{times} 
\usepackage{units} 
\def\etal{$\it{et~al.}$} 
\def\d{\partial \mu/\partial n}

\begin{document} 

\title{Measurement of the electronic compressibility of bilayer graphene} 

\author{E. A. Henriksen} 
\altaffiliation{Electronic address: erikku@caltech.edu} 
\affiliation{Condensed Matter Physics, California Institute of Technology, Pasadena, CA 91125}  

\author{J. P. Eisenstein} 
\affiliation{Condensed Matter Physics, California Institute of Technology, Pasadena, CA 91125}  
\begin{abstract}
We present measurements of the electronic compressibility, $K$, of bilayer graphene in both zero and finite magnetic fields up to 14 T, and as a function of both the carrier density and electric field perpendicular to the graphene sheet.  The low energy hyperbolic band structure of bilayer graphene is clearly revealed in the data, as well as a sizable asymmetry between the conduction and valence bands.  A sharp increase in $K^{-1}$ near zero density is observed for increasing electric field strength, signaling the controlled opening of a gap between these bands.  At high magnetic fields, broad Landau level (LL) oscillations are observed, directly revealing the doubled degeneracy of the lowest LL and allowing for a determination of the disorder broadening of the levels.
\end{abstract}
\date{\today}

\maketitle  

The hyperbolic band structure predicted for bilayer graphene stands apart from the quadratic dispersions common to almost all semiconductor systems at low energies, and is equally distinguished from the linear dispersion of monolayer graphene \cite{review}.  The curvature near the band edges rapidly evolves toward a near-linear form with increasing energy \cite{mccann:086805}.  Accordingly, the single particle density of states (DOS) has an unusual form, rising linearly with increasing energy from a finite value at zero energy.  Owing to the unique, tunable band gap of bilayer graphene, the DOS can be sharply modified reflecting the changing band structure \cite{mccann:161403,castro:216802}.  Here we have used an electric field penetration technique to measure the compressibility, $K$, of the two-dimensional electronic system in bilayer graphene, and thereby observe and explore changes in the DOS as a function of applied electric and magnetic fields \cite{eisenstein:1760}.  Thermodynamic quantities such as the compressibility are sensitive to both localized and extended electronic states, and are known to be an effective tool for investigating many-body effects that arise in interacting systems and have recently been predicted to occur in bilayer graphene \cite{eisenstein:1760,kusminskiy:106805,abergel:081408,borghi:2156}.  Our experimental results demonstrate that, contrary to the usual parabolic approximation made at low energies, the hyperbolic nature of the band structure must be accounted for to the lowest carrier densities measured.  Moreover, a clear asymmetry between the conduction and valence bands is revealed, implying a reduced DOS in the conduction band relative to the valence band.  When a strong electric field is applied perpendicular to the graphene sheet, a rising peak at zero density is consistent with the opening of a band gap.  At high magnetic fields, $B$, strong oscillations in the compressibility signal arise due to the formation of Landau levels (LLs), which are considerably broadened by disorder.

\begin{figure}[b]
\includegraphics[width=\columnwidth]{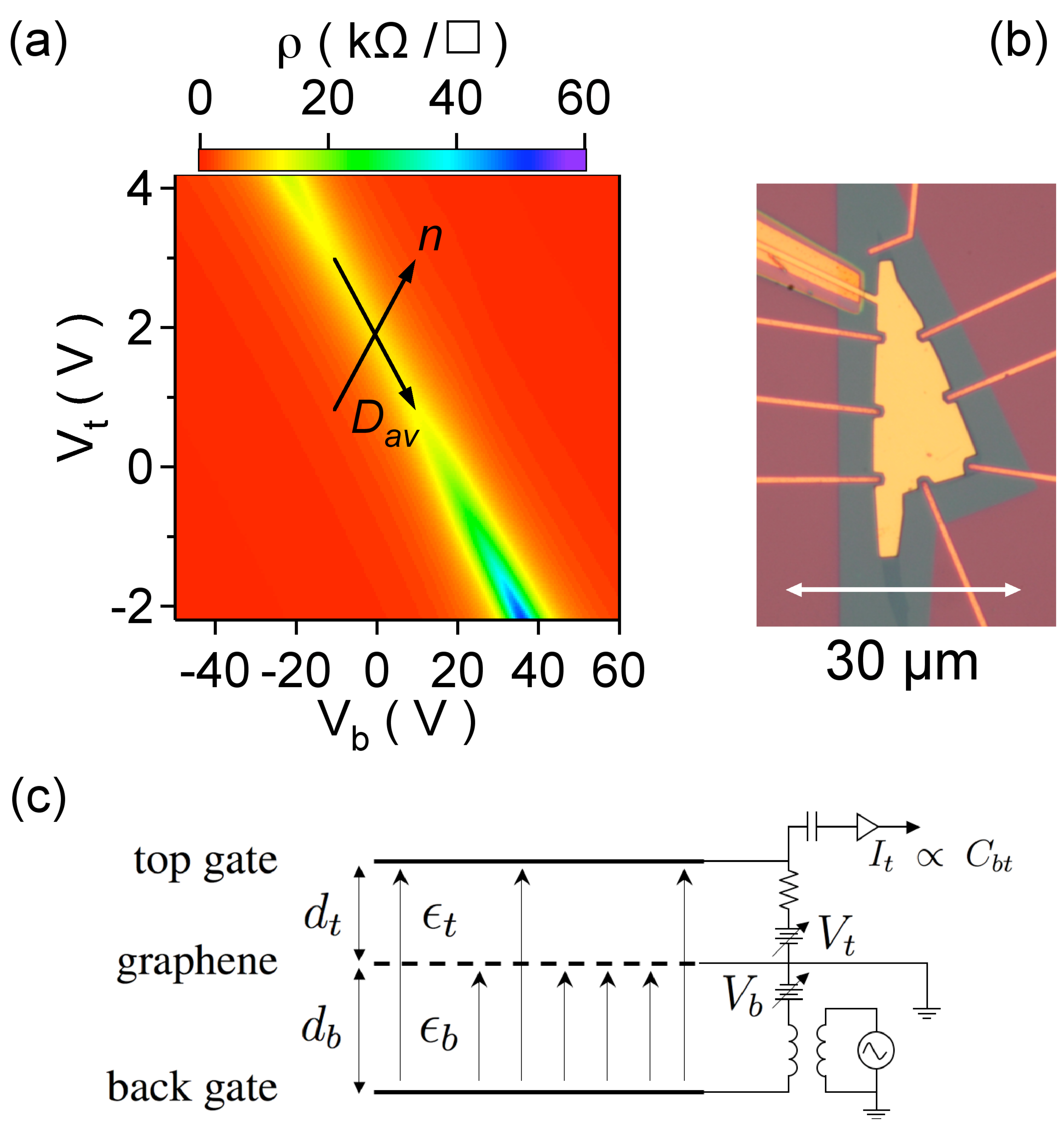} 
\caption{(Color online) (a) Sheet resistivity of the bilayer graphene device vs. top and back gate voltages $V_t$ and $V_b$.  The arrows indicate the orthogonal axes of the carrier density, $n$, and the displacement electric field perpendicular to the sheet, $D_{av}$.  (b) Optical microscope image of the device.  (c) Schematic of the electric field penetration measurement.} 
\end{figure}  

Bilayer graphene is composed of two monolayer graphene sheets in AB (Bernal) stacking.  To first order in the tight-binding (TB) approximation, the electronic dispersion of pristine, ungated bilayers consists of two nested hyperbolae in each of the conduction and valence bands, occurring at both the {\bf K} and {\bf K$^{\prime}$} points of the Brillouin zone \cite{review,mccann:086805}.  The two lowest energy bands meet at $k=0$, while the two higher bands are split off by a large interlayer hopping energy of about $\gamma_1=0.4$ eV.  This simple yet highly unusual dispersion can be radically altered by the application of an electric field normal to the plane, which breaks inversion symmetry and results in the opening of a {\it tunable} band gap between the lowest bands \cite{mccann:161403,castro:216802,oostinga:151,zhang:820,li:037403,kuzmenko:165406}.  In a dual-gated field-effect transistor device, where a bilayer is sandwiched between two electrical gates, the Fermi level (and hence carrier density) and the size of the gap can be independently manipulated \cite{oostinga:151,zhang:820}.  Besides holding promise for future applications, this allows for investigation of physics depending on only one or the other quantity.  With the bilayer embedded in a dielectric medium, the density and gap are controlled by the back and top gate biases, $V_t$ and $V_b$.  Electrical gating of the graphene results in a carrier density $n = \alpha (\Delta V_b- \beta \Delta V_t)$, while the average value of the electric displacement field perpendicular to the sheet, $D_{av}=(D_b+D_t)/2=(\epsilon_b \Delta V_b/d_b-\epsilon_t \Delta V_t/d_t)/2$, generates the band gap.  Here $\Delta V_i = V_i-V_{i0}$, where $V_{i0}$ is an offset voltage from charge neutrality (CN) due to extrinsic dopants, while $\epsilon_i$ and $d_i$ are the dielectric constant and thickness of each gate insulation.  

\begin{figure*}[t]
\includegraphics[width=2.1\columnwidth]{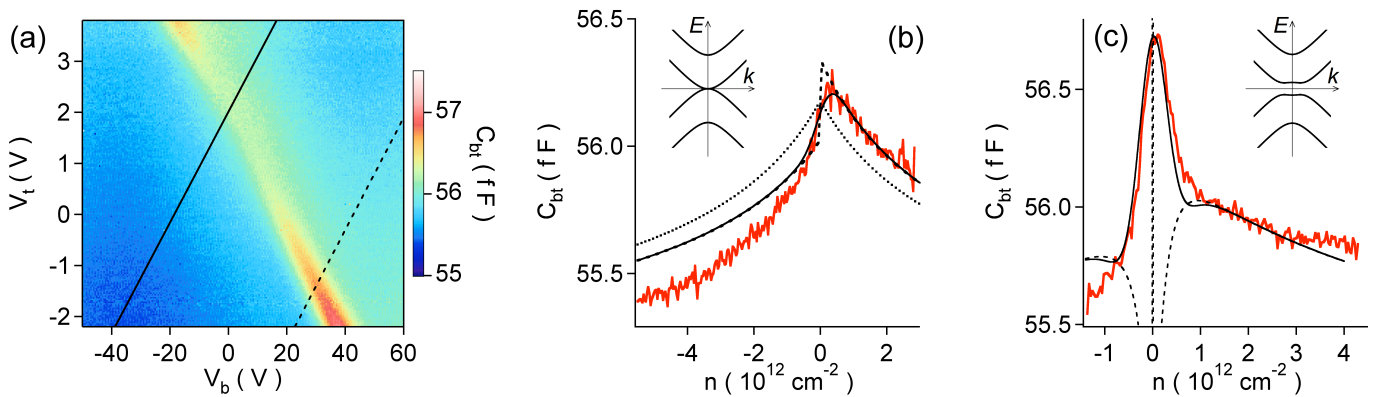} 
\caption{(Color online) (a) Measured back gate to top gate capacitance $C_{bt}$, proportional to $\d$, plotted vs. $V_t$ and $V_b$ at zero magnetic field.  The solid and dashed lines are profile cuts through the data while the insets schematically depict the band structure, for (b) $D_{av}=0$ (zero gap) and (c) $D_{av}=0.4$ V/nm (gapped), respectively.  The dotted, dashed, and solid lines in (b) and (c) are fits to the data described in the text.} 
\end{figure*}  

Our dual-gated devices begin with Scotch tape exfoliation of graphene bilayers onto Si/SiO$_2$ wafers having a 310 nm thick oxide.  Standard electron beam lithography and thin film deposition techniques are used to fabricate several Cr/Au contacts to the bilayer.  A layer of dilute 950K PMMA is then deposited and overexposed with an electron beam dosage of 24,000 $\mu$C/cm$^2$ to create a thin ($d_t=25-30$ nm), hardened dielectric layer of amorphous carbon over the graphene \cite{huard:236803,duan:135306}.  In the final fabrication step, a Cr/Au top gate is deposited, closely aligned to the contours of the graphene.  An optical microscope image of a typical finished device is shown in Fig. 1 (b).  The mobility of the sample used in this work is 2,900 cm$^2$/Vs, determined from the linear density dependence of the conductivity along the line $D_{av}=0$ \cite{adam:115436}.  All measurements were carried out at T=1.5 K.  

Electronic transport measurements were made using standard techniques and serve to identify the bilayer nature of the sample, as well as to calibrate $\alpha$ and $\beta$.  In Fig. 1 (a) we show a map of the sheet resistivity at zero magnetic field as a function of $V_b$ and $V_t$.  In this plot, the charge carriers are electrons to the right of the diagonal and holes to the left.  The peak resistivity occurs along the diagonal corresponding to CN; note the offset such that for $V_b=V_t=0$, the bilayer is hole-doped.  Along this CN diagonal, the minimum resistivity occurs at the saddle point located at $(V_{b0},V_{t0})=(-0.5,1.9)$, which we identify as where both $n$ and $D_{av}$, and hence the induced band gap, are zero.  Moving along the diagonal away from this point, the resistivity increases monotonically due to the opening of the band gap.  Together with the anomalous quantum Hall effect peculiar to bilayer graphene and observed in this sample, these transport measurements demonstrate that our sample is a graphene bilayer \cite{mccann:086805,novoselov:177}.  We establish the gate voltage to density calibration, $\alpha=7.1\times 10^{10}$ V$^{-1}$cm$^{-2}$, by counting oscillations in the compressibility data with changing LL filling factor, $\nu$, at finite magnetic field, as in Fig. 3 (a).  Here $\nu=n h/(g e B)$, where $g=4$ is the combined spin and valley degeneracies.  The slope of the CN diagonal in Fig. 1 (a) gives the ratio of the back to top gate capacitances per area, $1/\beta = - d_t \epsilon_b / (d_b \epsilon_t) = -0.11$.  Thus for SiO$_2$ with $\epsilon_b=3.9\epsilon_0$ we find $\epsilon_t=3.3\epsilon_0$ for the dielectric constant of the overexposed PMMA. 

The compressibility is measured using an electric field penetration technique schematically depicted in Fig. 1 (c) \cite{eisenstein:1760}.  In practice, we measure the capacitance of the back gate to the top gate, $C_{bt}$, where the electric field lines between the gates penetrate the grounded graphene sheet.  The finite DOS limits the ability of the graphene to screen the electric field lines, establishing a direct relationship between the measured quantity, $C_{bt}$, and the inverse DOS, $\d$, which is related to the electronic compressibility via $K^{-1} = n^2 \d$ where $\mu$ is the chemical potential.  For the device configuration used here, in the limit of an infinitely thin graphene sheet the measured capacitance is given by
\begin{equation}
C_{bt} = \frac{A_t d_g \epsilon_b \epsilon_t}{d_g (\epsilon_t d_b + \epsilon_b d_t) + \epsilon_0 d_b d_t}~+~C_s~.
\end{equation}
Here $A_t$ is the area of the top gate and $d_g = (\epsilon_0/e^2)\d$, with $-e$ the electron charge, is a convenient parameterization of $\d$ in units of length.  For typical experimental parameters, $d_g\approx2$ \AA.  Referring again to Fig. 1 (c), with the graphene held at ground we apply a small AC signal (typically 0.5 V at 100 Hz) to the back gate and record the resulting current, $I_t\propto C_{bt}$, flowing from the top gate.  Thus the quantity we are sensitive to is $\d$.  The parallel stray capacitance, $C_s$, arises from fringing fields both at the edge of the graphene flake, and between the top gate wiring and the entire Si substrate.  Shielding the top gate wiring with an on-chip ground plane, visible in Fig. 1 (b), and soldering the top gate wire to a small diameter coax cable held less than 1 mm above the device help to minimize the impact of $C_s$.  The quadrature signal in all measurements is small and constant across a range of frequencies $f=30-1000$ Hz, confirming the capacitive nature of the measurement.

Figure 2 (a) displays a map of the measured $C_{bt}$ at zero magnetic field, shown as a function of both $V_t$ and $V_b$.  Mimicking the sheet resistivity, the signal is strongest along the CN diagonal and monotonically declines with increasing hole or electron density.   For clarity, Fig. 2 (b) and (c) show two profile cuts through the data, taken along the lines of constant $D_{av}$ marked in Fig. 2 (a) and plotted as a function of the carrier density.  These mountain-like profiles evince a prime result of this work: since $C_{bt}\propto \d$, if any portion of the band structure were parabolic, a {\it constant} $C_{bt}$ vs. $n$ would result.  Such a parabolic approximation is often assumed for carrier densities $\lesssim4\times 10^{12}$ cm$^{-2}$.  Yet in each trace, the data monotonically and sharply decrease from a peak value near zero density. 

Insight into the decrease of $\d$ with $|n|$ can be gained from a minimal tight-binding model of bilayer graphene.  In the most basic TB model, the low energy bands of ungapped graphene give rise to a $\d$ that peaks at $n=0$ and falls off roughly as $1/n$ \cite{review,mccann:086805}.  Utilizing this model with the Fermi velocity set to $v_F=10^6$ m/s, along with Eq. 1 where $C_s$ is the only fitting parameter, we fit the $D_{av}=0$ trace with the dotted line in Fig. 2 (b).  We take $C_s=54.8$ fF, such that the remaining signal varies by roughly a factor of two over the range of densities measured.  Remarkably, this simple model reflects the general trend of the data, strongly suggesting the low energy band structure is in fact hyperbolic.  While anticipated theoretically, this low energy regime has not been clearly observed previously.  

A clear electron-hole asymmetry with higher $C_{bt}$ values for electron doping is visible in both Fig. 2 (b) and (c), which is absent from the transport data.  Both theory \cite{charlier:4579,zhang:235408,gava:165431} and experiment \cite{li:037403,kuzmenko:165406,malard:201401,henriksen:087403} expect that particle-hole symmetry is broken in bilayers leading to an asymmetry in the DOS.  Indeed, in the full TB model several terms can introduce such an asymmetry \cite{mucha:033001}; the most straightforward method is to include a hopping velocity, $v_4$, representing same-sublattice hopping between the two graphene sheets.  Inclusion of this term increases (decreases) the curvature of the conduction (valence) band, thereby reducing (increasing) the DOS so that in the absence of disorder, $\d$ undergoes a discontinuous jump between the two bands.  Repeating the fit with a value of $v_4=(0.6\pm0.2)\times10^5$ m/s yields the best result shown as the dashed curve in Fig. 2 (b).  This $v_4$ is close to the range $0.4-0.5\times 10^5$ m/s determined from spectroscopic experiments \cite{kuzmenko:165406,zhang:235408,malard:201401}.  

Generic to both mono- and bilayer graphene samples on a SiO$_2$ substrate are the so-called electron-hole ``puddles'' induced by charged impurities, which lead to inhomogeneous variations in the value of $n$ across the sample over a typical length scale of tens of nm \cite{martin:144,deshpande:243502}.  Such fluctuations are averaged over in our measurement due to the large area of the device, so we include this effect in our model by convolving the fits with a Gaussian and adjusting the variance, $\delta n$, until a best fit is found.  The solid line in Fig. 2 (b) is found for $\delta n = 3\times10^{11}$ cm$^{-2}$.  This $\delta n$ is consistent with other values in the literature for graphene-on-SiO$_2$ samples \cite{martin:144,deshpande:243502,yan:136804,hong:241415}.  

Overall, the fit is best at low densities, but becomes increasingly poor for higher hole densities suggesting our simple model is incomplete.  By including additional TB terms an improved fit may be achieved, but only at the cost of poorly constraining the fitting parameters.  While the relatively low mobility of our device implies that effects beyond the single-particle band structure are secondary, it is likely that $\d$ is impacted by screening and other many-body interactions not included in our model at this time.  In fact, recent calculations of the bilayer compressibility suggest interaction effects are not negligible \cite{kusminskiy:106805,abergel:081408,borghi:2156}.  In Ref.~\cite{kusminskiy:106805}, a negative divergence of $\d$ akin to that in parabolic 2D systems is predicted at very low densities.  Although we do not observe this phenomenon, the calculation does not account for the presence of disorder.  Alternatively, a recent calculation of interactions in the random phase approximation finds that $\d$ increases monotonically with decreasing density, and is always larger than the non-interacting value \cite{borghi:2156}.  While neither disorder nor an asymmetry effect are accounted for, the predicted increase in $\d$ may qualitatively account for the greater variation seen in the data compared to the TB model fits.

The profile cut for $D_{av}=0.4$ V/nm shown in Fig. 2 (c) exhibits a sharp peak centered at zero density, which we interpret as the opening of a band gap with concomitant changes in the DOS.  Extending the minimal model used above to include a gap suffices to describe this data trace.  We first solve the Hamiltonian for a gapped bilayer and find the associated $\d$, accounting for $v_4$ along with a band gap of size $\Delta$.  Then $C_{bt}$ is found using Eq. 1, and employing the same Gaussian smoothing over $\delta n$ as above we vary $\Delta$ until a best fit is obtained.  The dashed and solid lines in Fig. 2 (c) are the fits before and after smoothing.  For $D_{av}=0.4$ V/nm, the best fit is found with $\Delta=26$ meV.  This is considerably smaller than the 140 meV ``bare'' band gap associated with a field of this strength applied between the sheets of the bilayer.  Charge flow between the two layers in response to the external field will screen the bare gap \cite{mccann:161403,gava:165431,falkovsky:113413,fogler:5607}.  We note a gap of size $\Delta=26$ meV agrees to within uncertainty with that found in similar dual-gated devices at an equivalent displacement field $D_{av}$ as determined by IR spectroscopy \cite{zhang:820}.  Besides the density variations $\delta n$, the DOS will in general be broadened by disorder or trigonal warping which will diminish the band gap further and smooth over sharp features \cite{zhang:235408,nilsson:126801,raikh:195409}.  In the fitting so far we have ignored these additional sources of broadening, so that the value of $\delta n$ used to obtain good fits to the data should be taken as an upper limit.  

\begin{figure}[t]
\includegraphics[width=\columnwidth]{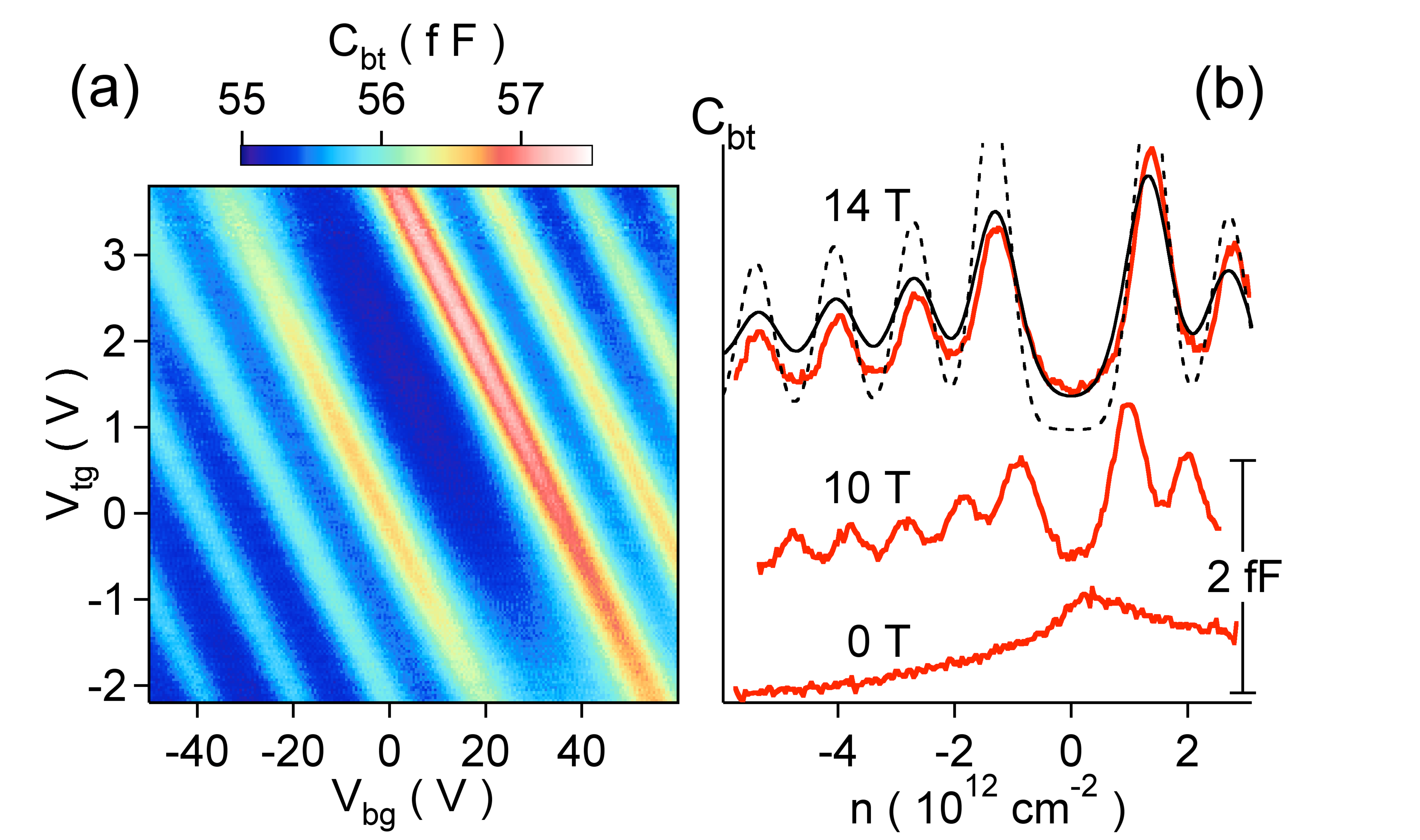} 
\caption{Color online) (a) Back gate to top gate capacitance $C_{bt}$ as a function of the top and back gate voltages, for $B = 14$ T.  (b) Profile cuts along the line $D_{av}=0$ for $B = 0, 10,$ and 14 T, offset for clarity.  Fits to the 14 T trace are for: density fluctuations only, $\delta n=3\times 10^{11}$ cm$^{-2}$ (dashed line); and $\delta n$ plus additional broadening of the LLs of width $\delta E = 11$ meV (solid line).} 
\end{figure} 

In strong magnetic fields the compressibility is greatly modified by Landau quantization.  In Fig. 3 (a) we display a map of $C_{bt}$ for a field $B = 14$ T, where the quantization alternately manifests as valleys, indicating compressible regions where a Landau level is partially filled, and ridges, when the Fermi level lies between a pair of LLs and the system is incompressible.  The central valley is twice as wide as the others due to the unusual bilayer LL structure which places two 4-fold degenerate LLs at zero energy \cite{mccann:086805}.  At 14 T, the gap opening has only a weak effect on the LL structure, apparent as an increase in $C_{bt}$ to the lower right of Fig. 3 (a), with an accompanying decrease in the peak heights of the incompressible ridges corresponding to LL filling factors $\nu=\pm4$.  The asymmetry observed at $B = 0$ T persists to high fields as stronger oscillations of $C_{bt}$ for electron vs. hole densities.  Qualitatively, the increased curvature of the conduction band results in greater LL spacings which, for a given LL broadening, should result in a smaller remnant DOS (and thus larger $C_{bt}$) in the gaps between LLs as compared to the valence band.  In Fig. 3 (b) cuts through the data for $B = 0, 10,$ and 14 T are plotted.  The evolution of the asymmetry and the increased range of $C_{bt}$ values with increasing $B$ are clearly visible.  We construct fits to the data at 14 T by first assuming the underlying LLs are $\delta$-functions centered at energies calculated within the minimal tight-binding picture plus a term accounting for the $v_4$ asymmetry \cite{pereira:115419,koshino}.  From this model DOS, $C_{bt}$ is calculated for the same values of $C_s$ and $\delta n$ as above.  The resulting fit, shown as the dashed line in Fig. 3 (b), greatly overestimates the amplitude of the $C_{bt}$ oscillations.  To improve on this, we have either to assume that $\delta n$ is nearly twice as large as the zero field value, or that an additional source of broadening is present.  While $\delta n$ will not vary greatly with $B$ field, disorder-broadening of LLs is expected in 2D systems including graphene \cite{shon:2421,zhu:056803,martin:669}.  The solid line fit in Fig. 3 (b) represents a calculation beginning with Lorentzian-broadened LLs all having equal width, $\delta E = 11$ meV.  This corresponds to a scattering time $\tau = 60$ fs, in good agreement with the transport scattering time $\tau_\mu=m^* \mu/e = 58$ fs, where $m^* = \gamma_1/2v_F^2$.  After further averaging over the same $\delta n$ as at zero field, a better agreement with the data is found.  Broadening with Gaussian-shaped LLs consistently produces a worse fit.  The asymmetry at high fields is not as well described by the zero field value of $v_4$.  As with the zero field fitting in Fig. 2 (b), the fits at high field match the data best at low carrier densities, and exhibit a similar discrepancy at high hole densities where the data fall below the fit.

The picture emerging from our data is clearly consistent with a hyperbolic band structure, across a wide density range relevant for many experiments on bilayer graphene.  Additionally, a sizeable conduction-valence band asymmetry is present.  Beyond the smoothing due to averaging over density fluctuations, in high magnetic fields additional broadening of the LLs is required to model our data.

{\it Note added in proof.} Recently, we became aware of similar measurements of the compressibility  of bilayer graphene by Young \etal \cite{young:5556}

We gratefully acknowledge conversations with J. Alicea, B. Chickering, M. Fogler, M. Koshino, A. H. MacDonald, E. McCann, D. Nandi, G. Refael, S. Das Sarma, and A. Young.  This work is supported by the NSF under grant No. DMR-0552270, and the DOE under grant No. DE-FG03-99ER45766.

\end{document}